\documentclass[aps,pra,preprintnumbers,showpacs,tightenlines]{revtex4}
\usepackage{amssymb}
\usepackage{amsmath}
\usepackage{graphicx}
\usepackage{epsfig}
\usepackage{subfigure}
\usepackage{amsfonts}
\usepackage{CJK}

\begin{document}

\title{Possible realization of entanglement, logical gates and quantum information
transfer with superconducting-quantum-interference-device qubits in cavity
QED}

\author{Chui-Ping Yang and Shih-I Chu}
\address{Department of Chemistry, University of Kansas, and Kansas Center\\
for Advanced Scientific Computing, Lawrence, Kansas 66045}
\author{Siyuan Han}
\address{Department of Physics and Astronomy, University of Kansas, Lawrence, Kansas\\
66045}

\date{\today}

\begin{abstract}
We present a scheme to achieve maximally entangled states, controlled
phase-shift gate and SWAP gate for two SQUID qubits (squbits), by placing
SQUIDs in a microwave cavity. We also show how to transfer quantum
information from one squbit to another. In this scheme, no transfer of
quantum information between the SQUIDs and the cavity is required, the
cavity field is only virtually excited and thus the requirement on the
quality factor of the cavity is greatly relaxed.
\end{abstract}

 \maketitle
\date{\today}

\begin{center}
{\bf I. INTRODUCTION}
\end{center}

A number of groups have proposed how to perform quantum logic using
superconducting devices such as Josephson junction circuits [1-3], Josephson
junctions [4-7], Cooper pair boxes [8-12] and superconducting quantum
interference devices (SQUIDs) [13-16]. These proposals play an important
role in building up superconducting quantum computers. In this paper, we
show a scheme for doing quantum logic with SQUID qubits in a microwave
cavity. The proposal merges ideas from the quantum manipulation with
atoms/ions in cavity QED [17-20]. The motivation for this scheme is
fivefold: (i) About six years ago, SQUIDs were proposed as candidates to
serve as the qubits for a superconducting quantum computer [21]. Recently,
people have presented many methods for demonstrating macroscopic coherence
of a SQUID [22-23] or performing a $single$ ``SQUID qubit'' logic operation
[13-16], but did not give much report on how to achieve quantum logic for
two SQUID qubits. As we know, the key ingredient in any quantum computation
is the two-qubit gate. The present scheme shows a way to implement
two-squbit quantum logic gates (here and below, ``squbit'' stands for
``SQUID qubit''). (ii) Compared with the other non-cavity SQUID-based
schemes where significant resources may be involved in coupling two distant
qubits, the present scheme may be simple as far as coupling qubits, since
the cavity mode acts as a ``bus'' and can mediate long-distance, fast
interaction between distant squbits. (iii) SQUIDs are sensitive to
environment. By placing SQUIDs into a superconducting cavity, decoherence
induced due to the external environment can be greatly suppressed, because
the cavity can be doubled as the magnetic shield for SQUIDs. (iv) It is
known that certain kinds of atoms/ions have a weak coupling with environment
and long decoherence time. Experiments have been made so far in the
cavity-atom/ions, which demonstrated the feasibility of small-scale quantum
computing. However, technically speaking, the cavity-SQUID scheme may be
preferable for demonstration purposes to the cavity-atom/ion proposals,
since SQUIDs can be easily embedded in a cavity while the latter requires
techniques for trapping atoms/ions. (v) Quantum computation based on
semiconductor quantum dots have been paid much interest, but recent reports
show that superconducting devices have relatively long decoherence time
[24,25] compared with quantum dots [26-30]. Decoherence time can reach the
order of 1$\mu s-5\mu s$ for superconducting devices [24,25]; while, for
quantum dots, typical decoherence times for ``the spin states of excess
conduction electrons'' and for ``charge states of excitons'' are,
respectively, on the order of 100$ns$ [26-28] and the order of 1$ns$
[28-30]).

This paper focuses on quantum logical gates (the controlled phase-shift gate
and the SWAP gate) of two squbits inside a cavity. The scheme doesn't
require any transfer of quantum information between the SQUID system and the
cavity, i.e., the cavity is only virtually excited. Thus, the cavity decay
is suppressed during the gate operations. In addition, we discuss how to
create maximally entangled states with two squbits and how to transfer
quantum information from one squbit to another.

The paper is organized as follows. In Sec. II, we introduce the Hamiltonian
of a SQUID coupled to a single-mode cavity field. In Sec. III, we consider a
SQUID driven by a classical microwave pulse. In Sec. IV, we discuss how to
achieve two-squbit maximally entangled states, logical gates and information
transfer from one squbit to another. A brief discussion of the experimental
issues and the concluding summary are given in Sec. V.

\begin{center}
{\bf II. SQUID COUPLED TO CAVITY FIELD}
\end{center}

Consider a system composed of a SQUID coupled to a single-mode cavity field
(assuming all other cavity modes are well decoupled to the three energy
levels of the SQUID). The Hamiltonian of the coupled system $H$ can be
written as a sum of the energies of the cavity field and the SQUID, plus a
term for the interaction energy:
\begin{equation}
H=H_c+H_s+H_I,
\end{equation}
where $H_c$, $H_s$ and $H_I$ are the Hamiltonian of the cavity field, the
Hamiltonian of the SQUID and the interaction energy, respectively.

The SQUIDs considered throughout this paper are rf SQUIDs each consisting of
Josephson tunnel junction enclosed by a superconducting loop (the size of an
rf SQUID is on the order of 10$\mu m-$100 $\mu m$). The Hamiltonian for an
rf SQUID (with junction capacitance $C$ and loop inductance $L$) can be
written in the usual form [31,32]
\begin{equation}
H_s=\frac{Q^2}{2C}+\frac{\left( \Phi -\Phi _x\right) ^2}{2L}-E_J\cos \left(
2\pi \frac \Phi {\Phi _0}\right) ,
\end{equation}
where $\Phi $, the magnetic flux threading the ring, and $Q$, the total
charge on the capacitor, are the conjugate variables of the system (with the
commutation relation $\left[ \Phi ,Q\right] =i\hbar $), $\Phi _x$ is the
static (or quasistatic) external flux applied to the ring, and $E_J$ $\equiv
I_c\Phi _0/2\pi $ is the Josephson coupling energy ($I_c$ is the critical
current of the junction and $\Phi _0=h/2e$ is the flux quantum).

\begin{figure}[tbp]
\includegraphics[bb=176 46 636 461, width=8.0 cm, clip]{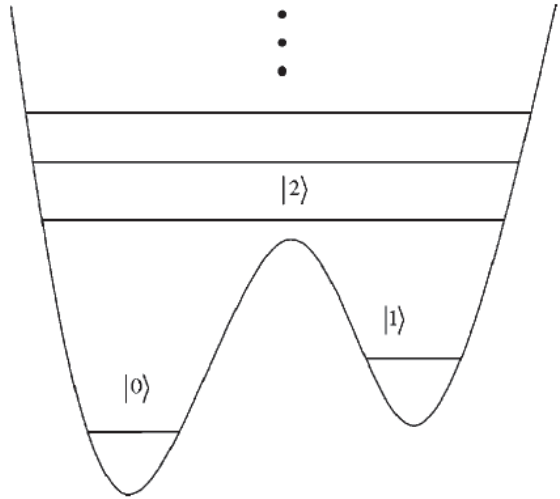} %
\vspace*{-0.08in}
\caption{Level diagram of a SQUID with the $\Lambda $-type three lowest
levels $\left| 0\right\rangle ,$ $\left| 1\right\rangle $ and $\left|
2\right\rangle .$}
\label{fig:1}
\end{figure}

The Hamiltonian of the single-mode cavity field can be written as
\begin{equation}
H_c=\hbar \omega _c\left( a^{+}a+\frac 12\right) ,
\end{equation}
where $a^{+}$ and $a$ are the creation and annihilation operators of the
cavity field; and $\omega _c$ is the frequency of the cavity field.

The cavity field and the SQUID ring are coupled together inductively with a
coupling energy given by
\begin{equation}
H_I=\lambda _c\left( \Phi -\Phi _x\right) \Phi _c,
\end{equation}
where $\lambda _c=-1/L$ is the coupling parameter linking the cavity field
to the SQUID ring; and $\Phi _c$ is the magnetic flux threading the ring,
which is generated by the magnetic component ${\bf B}\left( \stackrel{%
\rightarrow }{r},t\right) $ of the cavity field. The expression of $\Phi _c$
is given by
\begin{equation}
\Phi _c=\int_S{\bf B}\left( \stackrel{\rightarrow }{r},t\right) \cdot d{\bf S%
}
\end{equation}
( $S$ is any surface that is bounded by the ring, and $\stackrel{%
\rightharpoonup }{r}$ is the position vector of a point on $S$). ${\bf B}%
\left( \stackrel{\rightarrow }{r},t\right) $ takes the following form
\begin{equation}
{\bf B}\left( \stackrel{\rightarrow }{r},t\right) =\sqrt{\frac{\hbar \omega
_c}{2\mu _0}}\left[ a\left( t\right) +a^{+}\left( t\right) \right] {\bf B}%
\left( \stackrel{\rightarrow }{r}\right) ,
\end{equation}
where ${\bf B}\left( \stackrel{\rightarrow }{r}\right) $ is the magnetic
component of the normal mode of the cavity.

We denote $\left| n\right\rangle $ as the ($\Phi _x$-dependent) eigenstate
of $H_s$ with an eigenvalue $E_n$. Based on the completeness relation $%
\sum\limits_n$ $\left| n\right\rangle \left\langle n\right| =I,$ it follows
from (2) and (4) that
\begin{eqnarray}
H_s &=&\sum\limits_nE_n\left| n\right\rangle \left\langle n\right| ,
\nonumber \\
H_I &=&\sum\limits_n\left| n\right\rangle \left\langle n\right|
H_I\sum\limits_m\left| m\right\rangle \left\langle m\right| =\lambda _c\Phi
_c\sum\limits_{n,m}\left| n\right\rangle \left\langle n\right| \Phi -\Phi
_x\left| m\right\rangle \left\langle m\right| .
\end{eqnarray}
Let us consider the $\Lambda $-type three lowest levels of a SQUID, denoted
by $\left| 0\right\rangle ,$ $\left| 1\right\rangle $ and $\left|
2\right\rangle ,$ respectively (shown in Fig. 1). If the coupling of $\left|
0\right\rangle ,\left| 1\right\rangle $ and $\left| 2\right\rangle $ with
other levels via cavity modes is negligible, we have
\begin{equation}
H_s=E_0\left| 0\right\rangle \left\langle 0\right| +E_1\left| 1\right\rangle
\left\langle 1\right| +E_2\left| 2\right\rangle \left\langle 2\right|
\end{equation}
and
\begin{eqnarray}
H_I &=&\left( a+a^{+}\right) \left( g_{00}\left| 0\right\rangle \left\langle
0\right| +g_{11}\left| 1\right\rangle \left\langle 1\right| +g_{22}\left|
2\right\rangle \left\langle 2\right| \right)  \nonumber \\
&&\ \ +g_{01}a\left| 0\right\rangle \left\langle 1\right| +g_{12}a\left|
1\right\rangle \left\langle 2\right| +g_{02}a\left| 0\right\rangle
\left\langle 2\right|  \nonumber \\
&&\ \ +g_{10}a^{+}\left| 1\right\rangle \left\langle 0\right|
+g_{21}a^{+}\left| 2\right\rangle \left\langle 1\right| +g_{20}a^{+}\left|
2\right\rangle \left\langle 0\right|  \nonumber \\
&&\ \ +g_{01}a^{+}\left| 0\right\rangle \left\langle 1\right|
+g_{12}a^{+}\left| 1\right\rangle \left\langle 2\right| +g_{02}a^{+}\left|
0\right\rangle \left\langle 2\right|  \nonumber \\
&&\ \ +g_{10}a\left| 1\right\rangle \left\langle 0\right| +g_{21}a\left|
2\right\rangle \left\langle 1\right| +g_{20}a\left| 2\right\rangle
\left\langle 0\right| ,
\end{eqnarray}
where $g_{ii}=\lambda _c\sqrt{\frac{\hbar \omega _c}{2\mu _0}}\left(
\left\langle i\right| \Phi \left| i\right\rangle -\Phi _x\right) \widetilde{%
\Phi }_c$, $g_{ij}=\lambda _c\sqrt{\frac{\hbar \omega _c}{2\mu _0}}%
\left\langle i\right| \Phi \left| j\right\rangle \widetilde{\Phi }_c$ (here,
$\widetilde{\Phi }_c=$ $\int_S{\bf B}\left( \stackrel{\rightarrow }{r}%
\right) \cdot d{\bf S}$; $i,$ $j=0,1,2$ and $i\neq j$). For simplicity, we
will choose $g_{ij}=g_{ji}$ since eigenfunctions of $H_s$ can in general be
chosen to be real.

In the case when the cavity field is far-off resonant with the transition
between the levels $\left| 0\right\rangle $ and $\left| 1\right\rangle $ as
well as the transition between the levels $\left| 1\right\rangle $ and $%
\left| 2\right\rangle ,$ the Hamiltonian (9) reduces to
\begin{eqnarray}
H_I &=&\left( a+a^{+}\right) \left( g_{00}\left| 0\right\rangle \left\langle
0\right| +g_{11}\left| 1\right\rangle \left\langle 1\right| +g_{22}\left|
2\right\rangle \left\langle 2\right| \right)  \nonumber \\
&&+g_{02}a\left| 0\right\rangle \left\langle 2\right| +g_{20}a^{+}\left|
2\right\rangle \left\langle 0\right|  \nonumber \\
&&+g_{02}a^{+}\left| 0\right\rangle \left\langle 2\right| +g_{20}a\left|
2\right\rangle \left\langle 0\right| .
\end{eqnarray}
It follows from Eqs. (3), (8) and (10) that the interaction Hamiltonian in
the interaction picture is given by
\begin{eqnarray}
H_I &=&\left( e^{-i\omega _ct}a+e^{i\omega _ct}a^{+}\right) \left(
g_{00}\left| 0\right\rangle \left\langle 0\right| +g_{11}\left|
1\right\rangle \left\langle 1\right| +g_{22}\left| 2\right\rangle
\left\langle 2\right| \right)  \nonumber \\
&&\ +g_{02}e^{-i\left( \omega _c+\omega _{20}\right) t}a\left|
0\right\rangle \left\langle 2\right| +g_{20}e^{i\left( \omega _c+\omega
_{20}\right) t}a^{+}\left| 2\right\rangle \left\langle 0\right|  \nonumber \\
&&\ +g_{02}e^{i\left( \omega _c-\omega _{20}\right) t}a^{+}\left|
0\right\rangle \left\langle 2\right| +g_{20}e^{-i\left( \omega _c-\omega
_{20}\right) t}a\left| 2\right\rangle \left\langle 0\right| ,
\end{eqnarray}
where $\omega _{20}\equiv \left( E_2-E_0\right) /\hbar $ is the transition
frequency between the levels $\left| 0\right\rangle $ and $\left|
2\right\rangle .$

From (11) one can see that if the following condition is satisfied
\begin{equation}
\omega _c>>\Delta =\omega _c-\omega _{20},
\end{equation}
i.e., the cavity field frequency is much larger than the detuning from the
transition frequency between the levels $\left| 0\right\rangle $ and $\left|
2\right\rangle $, we can discard the rapidly oscillating terms in the
Hamiltonian (11) (i.e., the rotating-wave approximation). Thus, the final
effective interaction Hamiltonian (in the interaction picture) has the form
\begin{equation}
H_I=g_{02}\left[ e^{i\left( \omega -\omega _{20}\right) t}a^{+}\left|
0\right\rangle \left\langle 2\right| +e^{-i\left( \omega -\omega
_{20}\right) t}a\left| 2\right\rangle \left\langle 0\right| \right] ,
\end{equation}
where $g_{02}$ is the coupling constant between the SQUID and the cavity
field, corresponding to the transitions between $\left| 0\right\rangle $ and
$\left| 2\right\rangle .$

\begin{center}
{\bf III. SQUID DRIVEN BY A MICROWAVE PULSE}
\end{center}

Now, let's consider a SQUID driven by a classical microwave pulse (without
cavity). In the following, the SQUID is still treated quantum mechanically,
while the microwave pulse is treated classically. The Hamiltonian $H$ for
the coupled system can be written as
\begin{equation}
H=H_s+H_I,
\end{equation}
where $H_s$ and $H_I$ are the Hamiltonian (2) for the SQUID and the
interaction energy (between the SQUID and the microwave pulse),
respectively. The expression of $H_I$ is given by
\begin{equation}
H_I=\lambda _{\mu w}\left( \Phi -\Phi _x\right) \Phi _{\mu w},
\end{equation}
where $\lambda _{\mu w}=-1/L$ is a coupling coefficient linking the
microwave field to the SQUID ring; $\Phi _{\mu w}$ is the magnetic flux
threading the ring, which is generated by the magnetic component ${\bf B}%
^{\prime }\left( \stackrel{\rightarrow }{r},t\right) ={\bf B}^{\prime
}\left( \stackrel{\rightarrow }{r}\right) $ cos $\omega _{\mu w}t$ of the
microwave pulse, and has the following form
\begin{eqnarray}
\ \Phi _{\mu w} &=&\int_S{\bf B}^{\prime }\left( \stackrel{\rightarrow }{r}%
,t\right) \cdot d{\bf S}  \nonumber \\
\ &\equiv &\widetilde{\Phi }_{\mu w}\cos \omega _{\mu w}t
\end{eqnarray}
(here, $\widetilde{\Phi }_{\mu w}=\int_S{\bf B}^{\prime }\left( \stackrel{%
\rightarrow }{r}\right) \cdot d{\bf S,}$ the notations of $S$ and $\stackrel{%
\rightarrow }{r}$ are the same as described before, and $\omega _{\mu w}$ is
the frequency of the microwave pulse). Suppose that the microwave pulse is
resonant with the transition between the levels $\left| 0\right\rangle $ and
$\left| 2\right\rangle .$ Using the above procedures, the interaction
Hamiltonian in the interaction picture is then
\begin{eqnarray}
H_I &=&\Omega _{00}\left( e^{i\omega _{\mu w}t}+e^{-i\omega _{\mu w}}\right)
\left| 0\right\rangle \left\langle 0\right|  \nonumber \\
&&\ \ \ +\Omega _{22}\left( e^{i\omega _{\mu w}t}+e^{-i\omega _{\mu
w}}\right) \left| 2\right\rangle \left\langle 2\right|  \nonumber \\
&&\ \ \ +\Omega _{02}\left[ e^{-i\left( \omega _{\mu w}+\omega _{20}\right)
t}+e^{i\left( \omega _{\mu w}-\omega _{20}\right) t}\right] \left|
0\right\rangle \left\langle 2\right|  \nonumber \\
&&\ \ \ +\Omega _{20}\left[ e^{i\left( \omega _{\mu w}+\omega _{20}\right)
t}+e^{-i\left( \omega _{\mu w}-\omega _{20}\right) t}\right] \left|
2\right\rangle \left\langle 0\right| ,
\end{eqnarray}
where $\Omega _{ii}=\lambda _{\mu w}\left( \left\langle i\right| \Phi \left|
i\right\rangle -\Phi _x\right) \widetilde{\Phi }_{\mu w},$ $\Omega
_{ij}=\lambda _{\mu w}\left\langle i\right| \Phi \left| j\right\rangle
\widetilde{\Phi }_{\mu w}$ and $\Omega _{ij}=\Omega _{ji}$ ( $i,j=0,2$ and $%
i\neq j$). In the case of resonance ($\omega _{\mu w}=\omega _{20}$) and
under the rotating-wave approximation, the interaction Hamiltonian (17)
reduces to
\begin{equation}
H_I=\Omega _{02}\left( \left| 0\right\rangle \left\langle 2\right| +\left|
2\right\rangle \left\langle 0\right| \right) ,
\end{equation}
where $\Omega _{02}$ is the frequency of the Rabi oscillation between the
levels $\left| 0\right\rangle $ and $\left| 2\right\rangle .$ Based on (18),
it is easy to get the following state rotation
\begin{eqnarray}
\left| 0\right\rangle &\rightarrow &\cos \Omega _{02}t\left| 0\right\rangle
-i\sin \Omega _{02}t\left| 2\right\rangle ,  \nonumber \\
\left| 2\right\rangle &\rightarrow &-i\sin \Omega _{02}t\left|
0\right\rangle +\cos \Omega _{02}t\left| 2\right\rangle .
\end{eqnarray}
Similarly, when the microwave pulse frequency is tuned with the transition
frequency $\omega _{21}\equiv \left( E_2-E_1\right) /\hbar $ between the
levels $\left| 1\right\rangle $ and $\left| 2\right\rangle $, we have
\begin{equation}
H_I=\Omega _{12}\left( \left| 1\right\rangle \left\langle 2\right| +\left|
2\right\rangle \left\langle 1\right| \right) .
\end{equation}
Comparing $\left| 1\right\rangle $ and $\left| 2\right\rangle $ of Eq. (20)
with $\left| 0\right\rangle $ and $\left| 2\right\rangle $ of Eq. (18)
respectively, it is clear that we have
\begin{eqnarray}
\left| 1\right\rangle &\rightarrow &\cos \Omega _{12}t\left| 1\right\rangle
-i\sin \Omega _{12}t\left| 2\right\rangle ,  \nonumber \\
\left| 2\right\rangle &\rightarrow &-i\sin \Omega _{12}t\left|
1\right\rangle +\cos \Omega _{12}t\left| 2\right\rangle ,
\end{eqnarray}
where $\Omega _{12}=\lambda _{\mu w}\left\langle 1\right| \Phi \left|
2\right\rangle \widetilde{\Phi }_{\mu w}$ is the Rabi frequency between the
levels $\left| 1\right\rangle $ and $\left| 2\right\rangle .$

Finally, for the two-dimensional Hilbert space made of $\left|
0\right\rangle $ and $\left| 1\right\rangle ,$ an arbitrary rotation
\begin{eqnarray}
\left| 0\right\rangle &\rightarrow &\cos \Omega _{01}t\left| 0\right\rangle
-i\sin \Omega _{01}t\left| 1\right\rangle ,  \nonumber \\
\left| 1\right\rangle &\rightarrow &-i\sin \Omega _{01}t\left|
0\right\rangle +\cos \Omega _{01}t\left| 1\right\rangle ,
\end{eqnarray}
(where $\Omega _{01}=\lambda _{\mu w}\left\langle 0\right| \Phi \left|
1\right\rangle \widetilde{\Phi }_{\mu w}$) can be implemented if the
microwave frequency is tuned with the transition frequency $\omega
_{10}\equiv \left( E_1-E_0\right) /\hbar $ between the levels $\left|
0\right\rangle $ and $\left| 1\right\rangle .$ In the following discussions,
this rotation will not be employed, since it requires very long gate time
due to the barrier between the levels $\left| 0\right\rangle $ and $\left|
1\right\rangle $ [15].

\begin{figure}[tbp]
\includegraphics[bb=131 138 692 393, width=8.0 cm, clip]{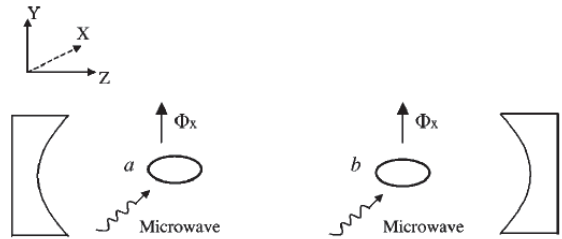} %
\vspace*{-0.08in}
\caption{Schematic illustration of two SQUIDs ($a,$ $b$) coupled to a
single-mode cavity field and manipulated by microwave pulses. The two SQUIDs
are placed along the cavity axis (i.e., the $Z$ axis). The microwave pulses
propagate in the $X$-$Z$ plane (parallel to the surface of the SQUID ring),
with the magnetic field component perpendicular to the surface of the SQUID
ring.}
\label{fig:2}
\end{figure}

\begin{center}
{\bf IV. ENTANGLEMENT, LOGICAL GATE, AND INFORMATION TRANSFER }
\end{center}

In this section, we consider two identical SQUIDs $a$ and $b$ coupled to a
single-mode microwave cavity (Fig. 2). The separation of the two SQUIDs is
assumed to be much larger than the linear dimension of each SQUID ring in
such a way that the interaction between the two SQUIDs is negligible. Also,
suppose that the coupling of each SQUID to the cavity field is the same
(this can be readily obtained by setting the two SQUIDs on two different
places $\stackrel{\rightarrow }{r}_1$ and $\stackrel{\rightarrow }{r}_2$ of
the cavity axis where the cavity-field magnetic components ${\bf B}\left(
\stackrel{\rightarrow }{r}_1,t\right) $ and ${\bf B}\left( \stackrel{%
\rightarrow }{r}_2,t\right) $ are the same). If the above assumption
applies, i.e., for each SQUID the coupling of the three lowest levels $%
\left| 0\right\rangle ,\left| 1\right\rangle $ and $\left| 2\right\rangle $
with other levels via cavity modes is negligible and the cavity field is
far-off resonant with the transition between the levels $\left|
0\right\rangle $ and $\left| 1\right\rangle $ as well as the transition
between the levels $\left| 1\right\rangle $ and $\left| 2\right\rangle ,$ it
is obvious that based on equation (13), the interaction Hamiltonian between
the two SQUIDs and the cavity field in the interaction picture can be
written as
\begin{equation}
H_I=g_{02}\sum_{m=a,b}\left( e^{-i\left( \omega _c-\omega _{20}\right)
t}a\left| 2\right\rangle _m\left\langle 0\right| +e^{i\left( \omega
_c-\omega _{20}\right) t}a^{+}\left| 0\right\rangle _m\left\langle 2\right|
\right) ,
\end{equation}
where the subscript $m$ represents SQUID $a$ or $b.$ In the case of $\omega
_c-\omega _{20}>>g_{02},$ i.e., the detuning between the transition
frequency (for the levels $\left| 0\right\rangle $ and $\left|
2\right\rangle $) and the cavity field frequency is much larger than the
corresponding coupling constant, there is no energy exchange between the
SQUIDs and the cavity field. The effective Hamiltonian is then given by
[33-34]
\begin{equation}
H=\gamma \left[ \sum\limits_{m=a,b}\left( \left| 2\right\rangle
_m\left\langle 2\right| aa^{+}-\left| 0\right\rangle _m\left\langle 0\right|
a^{+}a\right) +\left| 2\right\rangle _a\left\langle 0\right| \otimes \left|
0\right\rangle _b\left\langle 2\right| +\left| 0\right\rangle _a\left\langle
2\right| \otimes \left| 2\right\rangle _b\left\langle 0\right| \right] ,
\end{equation}
where $\gamma =g_{02}^2/\left( \omega -\omega _{20}\right) $. The first and
second terms of (24) describe the photon-number dependent Stark shifts,
while the third and fourth terms describe the ``dipole'' coupling between
the two SQUIDs mediated by the cavity mode. If the cavity field is initially
in the vacuum state, the Hamiltonian (24) reduces to
\begin{equation}
H=\gamma \left[ \sum\limits_{m=a,b}\left| 2\right\rangle _m\left\langle
2\right| +\left| 2\right\rangle _a\left\langle 0\right| \otimes \left|
0\right\rangle _b\left\langle 2\right| +\left| 0\right\rangle _a\left\langle
2\right| \otimes \left| 2\right\rangle _b\left\langle 0\right| \right] .
\end{equation}
Note that the Hamiltonian (25) does not contain the operators of the cavity
field. Thus, only the state of the SQUID system undergoes an evolution under
the Hamiltonian (25), i.e., no quantum information transfer exists between
the SQUID system and the cavity field. Therefore, the cavity field is
virtually excited.

It is clear that the states $\left| 0\right\rangle _a\left| 0\right\rangle
_b $ and $\left| 0\right\rangle _a\left| 1\right\rangle _b$ are unaffected
under the Hamiltonian (25) during the SQUID-cavity interaction. From (25),
one can easily get the following state evolution
\begin{eqnarray}
\left| 2\right\rangle _a\left| 0\right\rangle _b &\rightarrow &e^{-i\gamma
t}\left[ \cos \left( \gamma t\right) \left| 2\right\rangle _a\left|
0\right\rangle _b-i\sin \left( \gamma t\right) \left| 0\right\rangle
_a\left| 2\right\rangle _b\right] ,  \nonumber \\
\left| 0\right\rangle _a\left| 2\right\rangle _b &\rightarrow &e^{-i\gamma
t}\left[ \cos \left( \gamma t\right) \left| 0\right\rangle _a\left|
2\right\rangle _b-i\sin \left( \gamma t\right) \left| 2\right\rangle
_a\left| 0\right\rangle _b\right] ,  \nonumber \\
\left| 2\right\rangle _a\left| 2\right\rangle _b &\rightarrow &e^{-i2\gamma
t}\left| 2\right\rangle _a\left| 2\right\rangle _b,  \nonumber \\
\left| 2\right\rangle _a\left| 1\right\rangle _b &\rightarrow &e^{-i\gamma
t}\left| 2\right\rangle _a\left| 1\right\rangle _b.
\end{eqnarray}
In the following, we will show that Eq. (26) can be used to create
entanglement, to perform logical gates and to implement quantum information
transfer.

The operations described in the rest of this paper, can be realized by means
of the following three-step state manipulation: (i) first, adjust the level
spacing of each SQUID so that the transition between any two levels is
far-off resonant with the cavity field (in this case, the interaction
between the SQUIDs and the cavity field is turned off since the interaction
Hamiltonian (25) $H\approx 0$); (ii) apply a resonant microwave pulse to one
of the SQUIDs so that the state of this SQUID undergoes a transformation;
(iii) finally, adjust the level spacing of each SQUID back to the original
configuration, i.e., only the transitions $\left| 1\right\rangle
\leftrightarrow $ $\left| 2\right\rangle $ and $\left| 0\right\rangle
\leftrightarrow $ $\left| 1\right\rangle $ are far-off resonant with the
cavity field so that the system will undergo an evolution under the
Hamiltonian (25). In the SQUID system, the level spacing can be easily
changed by adjusting the external flux $\Phi _x$ or the critical current $%
I_c $ (for variable barrier rf SQUIDs). To simplify our discussion, we call
this 3-step process ``ARA'' (shown in Fig. 3).

\begin{figure}[tbp]
\includegraphics[bb=113 5 687 502, width=8.0 cm, clip]{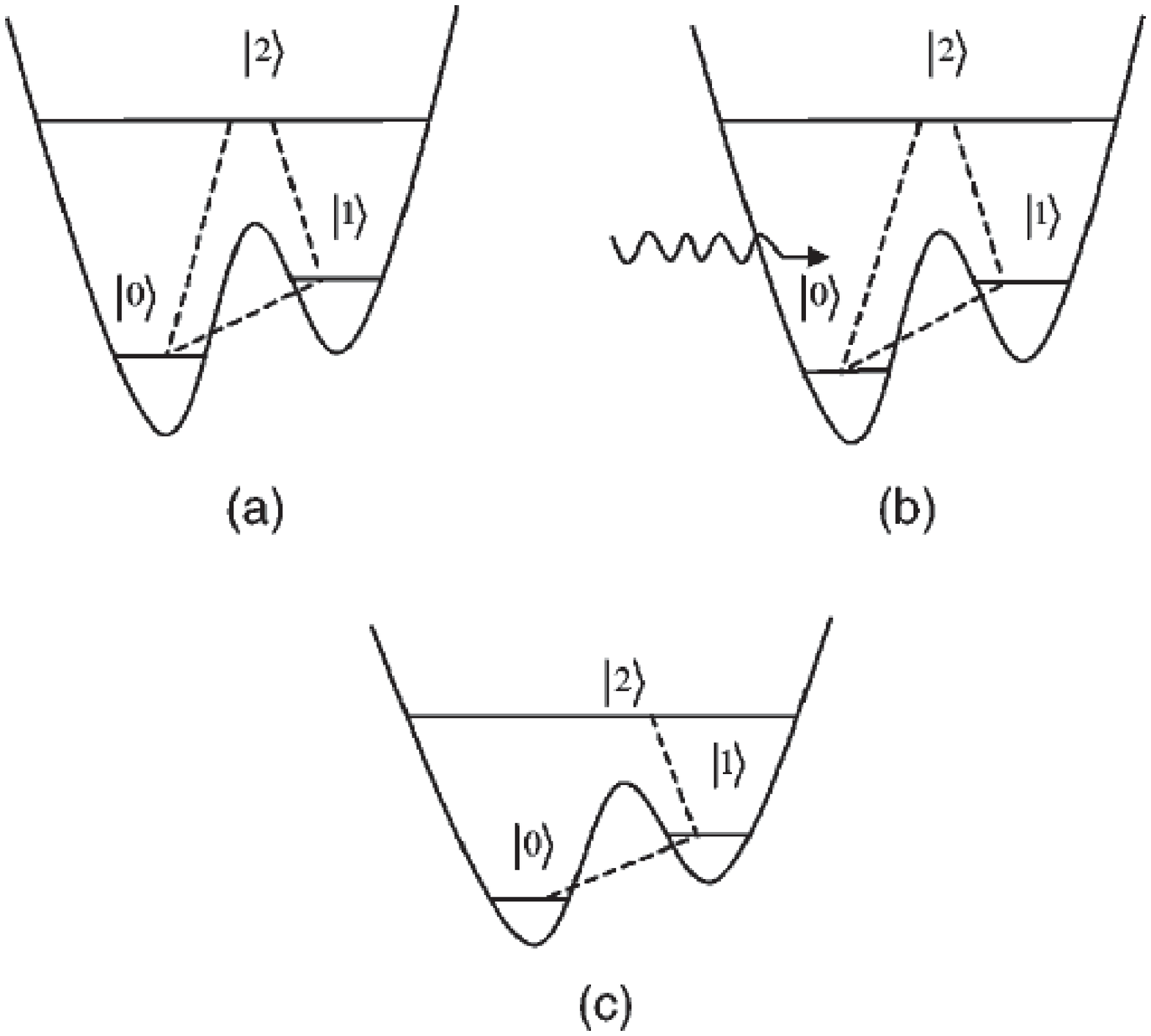} %
\vspace*{-0.08in}
\caption{Illustration of ARA. (i) the reduced level structure for each SQUID
after adjusting the level spacings; (ii) a microwave pulse with $\omega
_{\mu w}\equiv \omega _{20}$ or $\omega _{21}$ being applied to the SQUID $a$
or the SQUID $b$; (iii) the reduced level structure for each SQUID after
adjusting the level spacings back to that of before step (i). In Fig. 3 (i),
(ii) and (iii), the transition between levels linked by a dashed line is
far-off resonant with the cavity field.}
\label{fig:3}
\end{figure}

\begin{center}
{\bf A. generation of entanglement}
\end{center}

Entanglement is considered to be one of the most profound features of
quantum mechanics. An entangled state of a system consisting of two
subsystems cannot be described as a product of the quantum states of the two
subsystems. In this sense, the entangled system is considered inseparable
[35]. Recently, there has been much interest in practical applications of
entangled states in quantum computation, quantum cryptography, quantum
teleportation and so on [36-39]. Experimental realizations of entangled
states with up to four photons [40], up to four trapped ions [41] or two
atoms in microwave cavity QED [42] have been reported.

Assume that two SQUIDs are initially in the states $\left| 0\right\rangle _a$
and $\left| 0\right\rangle _b.$ In order to prepare the two squbits in the
maximally entangled state , we apply a ARA process in which a $\pi $-
microwave pulse (2$\Omega _{02}t=\pi ,$ where $t$ is the pulse duration),
resonant with the transition $\left| 0\right\rangle _a\leftrightarrow $ $%
\left| 2\right\rangle _a,$ is applied to the SQUID $a$. In this way, we
obtain the transformation $\left| 0\right\rangle _a\rightarrow -i\left|
2\right\rangle _a$, i.e., the state $\left| 0\right\rangle _a\left|
0\right\rangle _b$ becomes $-i\left| 2\right\rangle _a\left| 0\right\rangle
_b.$ After this ARA, let the state of the SQUID system evolve under the
Hamiltonian (25). From (26), one can see that after an interaction time $\pi
/\left( 4\gamma \right) $, the two SQUIDs will be in the maximally entangled
state
\begin{equation}
\left| \psi \right\rangle =-\frac 1{\sqrt{2}}\left( \left| 0\right\rangle
_a\left| 2\right\rangle _b+i\left| 2\right\rangle _a\left| 0\right\rangle
_b\right) ,
\end{equation}
where the common phase factor $e^{-i\pi /4}$ has been omitted. Note that the
rate of energy relaxation of the level $\left| 1\right\rangle $ is much
smaller than that of the level $\left| 2\right\rangle $ because of the
barrier between the levels $\left| 0\right\rangle $ and $\left|
1\right\rangle $ of the SQUIDs. Hence, to reduce decoherence, the state (27)
is transformed into
\begin{equation}
\left| \psi \right\rangle =\frac 1{\sqrt{2}}\left( i\left| 0\right\rangle
_a\left| 1\right\rangle _b-\left| 1\right\rangle _a\left| 0\right\rangle
_b\right)
\end{equation}
by applying a second ARA, in which each SQUID interacts with a $\pi $-
microwave pulse (resonant with $\omega _{21}$), resulting in the
transformation $\left| 2\right\rangle \rightarrow -i\left| 1\right\rangle $
for each SQUID. The prepared state (28) is a maximally entangled state of
two squbits $a$ and $b$ (here and in the following, the two orthogonal
states of a squbit are denoted by the two lowest energy states $\left|
0\right\rangle $ and $\left| 1\right\rangle $).

\begin{center}
{\bf B. controlled phase-shift gate }
\end{center}

Suppose that squbit $a$ is a control bit and squbit $b$ is a target bit. The
CPS gate can be realized in three steps:

Step (i): Apply a ARA in which a $\pi $-pulse with $\omega _{\mu w}=\omega
_{21}$ is applied to SQUID $a,$ resulting in the transformation $\left|
1\right\rangle _a\rightarrow -i\left| 2\right\rangle _a$.

Step (ii): After the ARA, let the state of the two SQUIDs undergo an
evolution for an interaction time $\pi /\gamma $ under the Hamiltonian (25).

Step (iii): Apply a ARA again in which a 3$\pi $-pulse with $\omega _{\mu
w}=\omega _{21}$ is applied to SQUID $a$, resulting in the transformation $%
\left| 2\right\rangle _a\rightarrow i\left| 1\right\rangle _a$.

The states of the 2-SQUID system after each step of the three
transformations are summarized in the following table:

\begin{equation}
\begin{array}{c}
\left| 0\right\rangle _a\left| 0\right\rangle _b \\
\left| 0\right\rangle _a\left| 1\right\rangle _b \\
\left| 1\right\rangle _a\left| 0\right\rangle _b \\
\left| 1\right\rangle _a\left| 1\right\rangle _b
\end{array}
\stackrel{\text{Step (i)}}{\rightarrow }
\begin{array}{c}
\left| 0\right\rangle _a\left| 0\right\rangle _b \\
\left| 0\right\rangle _a\left| 1\right\rangle _b \\
-i\left| 2\right\rangle _a\left| 0\right\rangle _b \\
-i\left| 2\right\rangle _a\left| 1\right\rangle _b
\end{array}
\stackrel{\text{Step (ii)}}{\rightarrow }
\begin{array}{c}
\left| 0\right\rangle _a\left| 0\right\rangle _b \\
\left| 0\right\rangle _a\left| 1\right\rangle _b \\
-i\left| 2\right\rangle _a\left| 0\right\rangle _b \\
i\left| 2\right\rangle _a\left| 1\right\rangle _b
\end{array}
\stackrel{\text{Step (iii)}}{\rightarrow }
\begin{array}{c}
\left| 0\right\rangle _a\left| 0\right\rangle _b \\
\left| 0\right\rangle _a\left| 1\right\rangle _b \\
\left| 1\right\rangle _a\left| 0\right\rangle _b \\
-\left| 1\right\rangle _a\left| 1\right\rangle _b
\end{array}
,
\end{equation}
which shows that a universal two-squbit CPS gate is realized.

A two-qubit CNOT gate can be obtained by combining a two-qubit CPS gate with
two single-qubit rotation gates [43]. Thus, applying the ARA procedures to
implement single-squbit rotating operations, together with the above CPS
gate operations, is sufficient to obtain the two-squbit CNOT gate.

\begin{center}
{\bf C. SWAP gate}
\end{center}

It is known that constructing a SWAP gate requires at least three CNOT gates
as follows [44]
\begin{eqnarray}
\left| i\right\rangle _a\left| j\right\rangle _b &\rightarrow &\left|
i\right\rangle _a\left| i\oplus j\right\rangle _b  \nonumber \\
\ &\rightarrow &\left| i\oplus \left( i\oplus j\right) \right\rangle
_a\left| i\oplus j\right\rangle _b=\left| j\right\rangle _a\left| i\oplus
j\right\rangle _b  \nonumber \\
\ &\rightarrow &\left| j\right\rangle _a\left| \left( i\oplus j\right)
\oplus j\right\rangle _b=\left| j\right\rangle _a\left| i\right\rangle _b,
\end{eqnarray}
where $i,j\in \left\{ 0,1\right\} $ and all additions are done modulo 2. As
described above, a CNOT can be realized with a CPS and two single-qubit
rotations. Since each two-squbit CPS gate requires three basic steps
described above, at least nine basic steps for three CPS gates, together
with six single-squbit rotation operations, are needed to implement a
two-squbit SWAP gate by using the above method. In the following discussion
we present a new way to perform a SWAP, which requires only five steps.

Step (i): apply a ARA in which each SQUID interacts with a $\pi $-pulse
(resonant with $\omega _{21}$)$,$ so that each SQUID undergoes the
transformation $\left| 1\right\rangle \rightarrow -i\left| 2\right\rangle $.

Step (ii): let the state of the SQUID system undergo an evolution for an
interaction time $\pi /\left( 2\gamma \right) $ under the Hamiltonian (25).

Step (iii): perform a ARA in which a 2$\pi $-pulse and a $\pi $-pulse,
resonant with $\omega _{21}$ of the SQUID $a$ and the SQUID $b$
respectively, are applied, resulting in transformations $\left|
2\right\rangle _a\rightarrow -\left| 2\right\rangle _a$ and $\left|
2\right\rangle _b\rightarrow -i\left| 1\right\rangle _b$.

Step (iv): let the state of the system undergo an evolution for an
interaction time $\pi /\gamma $ under the Hamiltonian (25).

Step (v): perform a ARA in which a 3$\pi $-pulse, resonant with $\omega
_{21},$ is applied to the SQUID $a$ so that it undergoes the transformation $%
\left| 2\right\rangle _a\rightarrow i\left| 1\right\rangle _a$ $.$

The states after each step of the above operations are listed below:
\begin{eqnarray}
&&
\begin{array}{c}
\left| 0\right\rangle _a\left| 0\right\rangle _b \\
\left| 0\right\rangle _a\left| 1\right\rangle _b \\
\left| 1\right\rangle _a\left| 0\right\rangle _b \\
\left| 1\right\rangle _a\left| 1\right\rangle _b
\end{array}
\stackrel{\text{Step (i)}}{\rightarrow }
\begin{array}{c}
\left| 0\right\rangle _a\left| 0\right\rangle _b \\
-i\left| 0\right\rangle _a\left| 2\right\rangle _b \\
-i\left| 2\right\rangle _a\left| 0\right\rangle _b \\
-\left| 2\right\rangle _a\left| 2\right\rangle _b
\end{array}
\stackrel{\text{Step (ii)}}{\rightarrow }
\begin{array}{c}
\left| 0\right\rangle _a\left| 0\right\rangle _b \\
i\left| 2\right\rangle _a\left| 0\right\rangle _b \\
i\left| 0\right\rangle _a\left| 2\right\rangle _b \\
\left| 2\right\rangle _a\left| 2\right\rangle _b
\end{array}
\stackrel{\text{Step (iii)}}{\rightarrow }
\begin{array}{c}
\left| 0\right\rangle _a\left| 0\right\rangle _b \\
-i\left| 2\right\rangle _a\left| 0\right\rangle _b \\
\left| 0\right\rangle _a\left| 1\right\rangle _b \\
i\left| 2\right\rangle _a\left| 1\right\rangle _b
\end{array}
\nonumber \\
&&\stackrel{\text{Step (iv)}}{\rightarrow }
\begin{array}{c}
\left| 0\right\rangle _a\left| 0\right\rangle _b \\
-i\left| 2\right\rangle _a\left| 0\right\rangle _b \\
\left| 0\right\rangle _a\left| 1\right\rangle _b \\
-i\left| 2\right\rangle _a\left| 1\right\rangle _b
\end{array}
\stackrel{\text{Step (v)}}{\rightarrow }
\begin{array}{c}
\left| 0\right\rangle _a\left| 0\right\rangle _b \\
\left| 1\right\rangle _a\left| 0\right\rangle _b \\
\left| 0\right\rangle _a\left| 1\right\rangle _b \\
\left| 1\right\rangle _a\left| 1\right\rangle _b
\end{array}
.
\end{eqnarray}
It is clear that the operations accomplish a two-squbit SWAP gate.

\begin{center}
{\bf D. transfer of information}
\end{center}

Recently, quantum teleportation [38] has been paid much interest because it
plays an important role in quantum information processing. It is also noted
that short-distance quantum teleportation can be applied to transport
quantum information inside a quantum computer [45]. It is well known that
transferring quantum information from one qubit to another requires a
minimum number of $three$ $qubits$ by using the standard teleportation
protocols [38,45]. In the following, we will present a different approach
for transferring quantum information from one squbit to another, by the use
of only two squbits.

Suppose that the squbit $a$ is the original carrier of quantum information,
which is in an arbitrary state $\alpha \left| 0\right\rangle +\beta \left|
1\right\rangle ;$ and we want to transfer this state from squbit $a$ to
squbit $b.$ To do this, the squbit $b$ is first prepared in the state $%
\left| 0\right\rangle $. The quantum state transfer between the two squbits
is described by
\begin{equation}
\left( \alpha \left| 0\right\rangle _a+\beta \left| 1\right\rangle _a\right)
\left| 0\right\rangle _b\rightarrow \left| 0\right\rangle _a\left( \alpha
\left| 0\right\rangle _b+\beta \left| 1\right\rangle _b\right) .
\end{equation}
From (32) one can see that this process can be done via a transformation
that satisfies the following truth table:
\begin{eqnarray}
\left| 0\right\rangle _a\left| 0\right\rangle _b &\rightarrow &\left|
0\right\rangle _a\left| 0\right\rangle _b,  \nonumber  \label{18} \\
\left| 1\right\rangle _a\left| 0\right\rangle _b &\rightarrow &\left|
0\right\rangle _a\left| 1\right\rangle _b,
\end{eqnarray}
which can be realized in three steps:

Step (i): perform a ARA in which a $\pi $-pulse ($\omega _{\mu w}=\omega
_{21}$) is applied to the SQUID $a,$ resulting in the transformation $\left|
1\right\rangle _a\rightarrow -i\left| 2\right\rangle _a$ $.$

Step (ii): let the state of the two SQUIDs undergo an evolution for an
interaction time $\pi /\left( 2\gamma \right) $ under the Hamiltonian (25).

Step (iii): perform a ARA in which a $\pi $-pulse ($\omega _{\mu w}=$ $%
\omega _{21}$) is applied to the SQUID $b,$ resulting in the transformation $%
\left| 2\right\rangle _b\rightarrow -i\left| 1\right\rangle _b$.

The truth table of the entire operation is summaried below:
\begin{equation}
\begin{array}{c}
\left| 0\right\rangle _a\left| 0\right\rangle _b \\
\left| 1\right\rangle _a\left| 0\right\rangle _b
\end{array}
\stackrel{\text{Step (i)}}{\rightarrow }
\begin{array}{c}
\left| 0\right\rangle _a\left| 0\right\rangle _b \\
-i\left| 2\right\rangle _a\left| 0\right\rangle _b
\end{array}
\stackrel{\text{Step (ii)}}{\rightarrow }
\begin{array}{c}
\left| 0\right\rangle _a\left| 0\right\rangle _b \\
i\left| 0\right\rangle _a\left| 2\right\rangle _b
\end{array}
\stackrel{\text{Step (iii)}}{\rightarrow }
\begin{array}{c}
\left| 0\right\rangle _a\left| 0\right\rangle _b \\
\left| 0\right\rangle _a\left| 1\right\rangle _b
\end{array}
.
\end{equation}
It is easy to verify that the operations described above achieve the desired
2-squbit teleportation (32).

From above descriptions, one can also see that in each ARA process, no
simultaneous $\left| 0\right\rangle \rightarrow \left| 2\right\rangle $ and $%
\left| 1\right\rangle \rightarrow \left| 2\right\rangle $ transitions are
required for each SQUID and hence it is unnecessary to have the microwave
pulses applied to two SQUIDs at the same time. Thus, it is sufficient to use
only one microwave source with fixed frequency $\omega _{\mu w},$ since the
transition frequency $\omega _{20}$ and $\omega _{21}$ of each SQUID can be
rapidly adjusted to meet the resonant condition ($\omega _{\mu w}=\omega
_{ij}$), and the microwave can be redirected from one SQUID to another.

\begin{center}
{\bf V. DISCUSSION AND CONCLUSION}
\end{center}

Some experimental matters may need to be addressed here. Firstly, the
required time $t_{op}$ for any gate operation (SWAP, CPS, CNOT etc.) should
be shorter than the energy relaxation time $t_r$ of the level $\left|
2\right\rangle $. The lifetime of the cavity mode is given by $T_c=Q/2\pi
\nu $ where $Q$ is the quality factor of the cavity and $\nu $ is the cavity
field frequency. In our scheme, the cavity has a probability $P\simeq $ $%
t_{op}/t_r$ of being excited during the operation. Thus the effective decay
time of the cavity is $T_c/P,$ which should be larger than the energy
relaxation time $t_r,$ i.e., the quality factor of the cavity should satisfy
$Q>>2\pi \nu t_{op}.$ The SQUIDs can be designed so that the level $\left|
2\right\rangle $ has a sufficiently long energy relaxation time and thus the
spontaneous decay of the SQUIDs is negligible during the operation. On the
other hand, we can also use a high-$Q$ cavity and reduce the operation time
by increasing the intensity of the microwave pulses and/or the coupling
constant $g_{02}$ (e.g., by varying the energy level structure of the
SQUIDs), so that the cavity dissipation is negligible during the operation.

For the sake of definitiveness, let us consider the SQUIDs described in Ref.
[15] for which the energy relaxation time $t_r$ of the level $\left|
2\right\rangle $ could exceed 1 $\mu s$ [24]$,$ the transition frequency $%
\nu _0$ between $\left| 0\right\rangle $ and $\left| 2\right\rangle $ is on
the order of $80$ GHz, and the typical gate time is $t_{op}\simeq 0.01t_r.$
Taking $t_r=1$ $\mu s,$ $\nu _0=80$ GHz and the detuning $\nu -\nu _0=$ 0.1
GHz, a simple calculation shows that the quality factor of the required
cavity should be greater than $5\times 10^3,$ which is readily available in
most laboratories. For instance, a superconducting cavity with a quality
factor $Q=10^8$ has been demonstrated by M. Brune et. al. [46].

It can be seen that the key element of the scheme is the ARA process. As
discussed previously, the realization of ARA requires rapid adjustments of
level spacings of SQUIDs. The applied microwave pulses are ensured to be
far-off resonant with the cavity field during each ARA because $\omega _{20}$
and $\omega _{21}$ are highly detuned from $\omega _c$. Thus, the use of the
microwave pulses does not change photon population in the cavity field. The
scheme presented here has the following advantages: (i) using only two
squbits (teleportation); (ii) faster (using 3-level gates) [15]; (iii) not
requiring very high-$Q$ microwave cavity; (iv) no need of changing the
microwave frequency $\omega _{\mu w}$ during the entire operation for all of
the gates described; (v) possibility of being extended to perform quantum
computing on lots of squbits inside a cavity (shown in Fig. 4) due to
long-distance coherent interaction between squbits mediated via the cavity
mode.

\begin{figure}[tbp]
\includegraphics[bb=143 188 644 340, width=8.0 cm, clip]{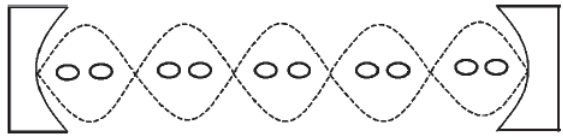} %
\vspace*{-0.08in}
\caption{Set up for quantum computing with many SQUIDs in a cavity. The
interaction between any two SQUIDs is mediated through a single-mode
standing-wave cavity field. During a logical gate operation on any two
chosen SQUIDs, all other SQUIDs can be decoupled, by adjusting the level
spacings so that the transition between any two levels of each other SQUID
is far-off resonant with the cavity field.}
\label{fig:4}
\end{figure}

Before we conclude, we should mention that the idea of coupling multiple
qubits globally with a resonant structure and tuning the individual qubits
to couple and decouple them from the resonator has been previously presented
for charge-based qubits [9]. Our scheme is much in the same spirit in the
sense of coupling and decoupling the individual qubits by manipulating their
Hamiltonians, but it is for a different system and it differs in the details
of both the qubits and the coupling structure. In our case, we consider a
system consisting of flux-based qubits (SQUIDs) coupled via a single-mode
microwave cavity field, while the system described in [9] comprises charge
qubits and a LC-oscillator mode in the circuit. The two logic states of a
qubit in our scheme are represented by the two lowest energy fluxoid states
of the SQUID, while the two logic states of a qubit in [9] are the two
charge states differing by one Cooper pair. More importantly, since the
scheme in [9] uses an inductor (which is a lumped circuit element) to couple
charge-based qubits, the frequency of the LC-oscillator mode, $\omega
_{LC}=1/\sqrt{NC_{qb}L},$ where $C_{qb}$ is the capacitance of each charge
qubit and $N$ is the number of qubits, decreases with the increase of the
number of qubits. Thus, the necessary condition for the coupling to work, $%
\hbar \omega _{LC}>>E_J,E_{ch},k_BT,$ where $E_J$ and $E_{ch}$ are the
energy scales of a charge qubit (for the detail, see [9]), becomes more
difficult to satisfy as the number of qubits increases. This problem does
not exist in our scheme since, to the first order, the frequency of the
cavity field is independent of the number of qubits. Therefore, in
principle, our scheme can be used to establish coupling among a large number
of qubits.

In summary, we have proposed a new scheme to create two-squbit maximally
entangled state and to implement two-squbit logical gates (SWAP, CPS and
CNOT) with the use of a microwave cavity. The method can also be used to
realize information transfer from one to another squbit (local
teleportation) with two, instead of three qubits. The method does not
require the transfer of quantum information between the cavity and the SQUID
system. The cavity is only virtually excited during the whole operation;
thus the requirement on the quality factor of the cavity is greatly relaxed.
The present proposal provides a new approach to quantum computing and
communication with superconducting qubits. To the best of our knowledge,
there has been no experimental demonstration of entanglement or logical
gates for two SQUIDs; and we hope that the proposed approach will stimulate
further theoretical and experimental activities.

\begin{center}
{\bf ACKNOWLEDGMENTS}
\end{center}

We thank Zhongyuan Zhou and Shi-Biao Zheng for many fruitful discussions and
Julio Gea-Banacloche for very useful comments. This work was partially
supported by National Science Foundation (EIA-0082499), and AFOSR
(F49620-01-1-0439), funded under the Department of Defense University
Research Initiative on Nanotechnology (DURINT) Program and by the ARDA.


\begin{references}
\bibitem{s1}  J. E. Mooij, T. P. Orlando, L. Levitov, L. Tian, C. H. van der
Wal, and S. Lloyd, Science {\bf 285}, 1036 (1999).

\bibitem{s2}  C. H. van der Wal, A. C. J. ter Haar, F. K. Wilhelm, R. N.
Schouten, C. J. P. M. Harmans, T. P. Orlando, S. Lloyd, and J. E. Mooij,
Science {\bf 290}, 773 (2000).

\bibitem{s3}  T. P. Orlando, J. E. Mooij, L. Tian, C. H. van der Wal, L.
Levitov, S. Lloyd, and J. J. Mazo, Phys. Rev. B {\bf 60}, 15398 (1999).

\bibitem{s4}  A. Shnirman, G. Sch\"on, and Z. Hermon, Phys. Rev. Lett. {\bf %
79}, 2371 (1997).

\bibitem{s5}  A. Blais, and A. M. Zagoskin, Phys. Rev. A {\bf 61}, 042308
(2000).

\bibitem{s6}  A. Steinbach, P. Joyez, A. Cottet, D. Esteve, M. H. Devoret,
M. E. Huber, and J. M. Martinis, Phys. Rev. Lett. {\bf 87}, 137003 (2001).

\bibitem{s7}  J. M. Martinis and R. L. Kautz, Phys. Rev. Lett. {\bf 63},
1507 (1989).

\bibitem{s8}  Y. Makhlin, G. Schoen, and A. Shnirman, Rev. of Mod. Phys.
{\bf 73}, 357 (2001).

\bibitem{s9}  Y. Makhlin, G. Schoen, and A. Shnirman, Nature {\bf 398}, 305
(1999).

\bibitem{s10}  Y. Nakamura, Y. Pashkin, and J. S. Tsai, Nature {\bf 398},
786 (1999).

\bibitem{s11}  W. Xiang-bin and M. Keiji, Phys. Rev. B {\bf 65}, 172508
(2002) (quant-ph/0104127).

\bibitem{s12}  W. Xiang-bin and M. Keiji, quant-ph/0105008.

\bibitem{s13}  J. R. Friedman, V. Patel, W. Chen, S. K. Tolpygo, and J. E.
Lukens, Nature {\bf 406}, 43 (2000).

\bibitem{s14}  X. Zhou, J. L. Habif, M. F. Bocko, and M. J. Feldman,
quant-ph/0102090.

\bibitem{s15}  Z. Zhou, Shih-I Chu and S. Han, Phys. Rev. B, {\bf 66},
054527 (2002).

\bibitem{s16}  P. Silvestrini and L. Stodolsky, cond-mat/0004472.

\bibitem{s17}  J. I. Cirac and P. Zoller, Phys. Rev. Lett. {\bf 74,} 4091
(1995).

\bibitem{s18}  T. Sleator and H. Weinfurter, Phys. Rev. Lett. {\bf 74,} 4087
(1995).

\bibitem{s19}  M. Brune, S. Haroche, J. M. Raimond, L. Davidovich, and N.
Zagury, Phys. Rev. A {\bf 45}, 5193 (1992).

\bibitem{s20}  Q. A.Turchette, C. J. Hood, W. Lange, H. Mabuchi, and H. J.
Kimble, Phys. Rev. Lett. {\bf 75}, 4710 (1995).

\bibitem{s21}  M. F. Bocko, A. M. Herr, and M. J. Feldman, IEEE Transactions
on Applied Superconductivity vol. {\bf 7}, no. 2, pt.3 3638-41 (1997).

\bibitem{s22}  R. C. Rey-de-Castro, M. F. Bocko, A. M. Herr, C. A. Mancini,
and M. J. Feldman, quant-ph/0102089

\bibitem{s23}  M. Crogan, S. Khlebnikov, and G. Sadiek, quant-ph/0105038.

\bibitem{s24}  Y. Yu, S. Han, X. Chu, S.-I. Chu, and Z. Wang, Science {\bf %
296}, 889 (2002).

\bibitem{s25}  D. Vion, A. Aassime, A.Cottet, P. Joyez, H. Pothier, C.
Urbina, D. Esteve and M. Devoret, Science {\bf 296}, 886 (2002).

\bibitem{s26}  J. M. Kikkawa and D. D. Awschalom, Phys. Rev. Lett. {\bf 80},
4313 (1998).

\bibitem{s27}  A. Imamoglu, D. D. Awschalom, G. Burkard, D. P. DiVincenzo,
D. Loss, M. Sherwin, and A. Small, Phys. Rev. Lett. {\bf 83}, 4204 (1999).

\bibitem{s28}  E. Pazy, E. Biolatti, T. Calarco, I. D'Amico, P. Zanardi, F.
Rossi, and P. Zoller, cond-mat/0109337.

\bibitem{s29}  M. Bayer and A. Forchel, Phys. Rev. B {\bf 65}, 041308(R)
(2002).

\bibitem{s30}  E. Biolatti {\it et} {\it al}$.$, Phys. Rev. Lett. 85, 5647
(2000); E. Biolatti {\it et} {\it al.}, Phys. Rev. B {\bf 65}, 075306 (2002).

\bibitem{s31}  S. Han, R. Rouse, and J. E. Lukens, Phys. Rev. Lett. {\bf 76}%
, 3404 (1996).

\bibitem{s32}  T. P. Spiller, T. D. Clark, R. J. Prance, and A. Widom, Prog.
Low Temp. Phys. {\bf 13}, 219 (1992).

\bibitem{s33}  S. B. Zheng and G. C. Guo, Phys. Rev. Lett. {\bf 85,} 2392
(2000); S. B. Zheng and G. C. Guo, Phys. Rev. A. {\bf 63,} 044302 (2001).
Zheng and Guo addressed how to realize quantum entanglement and the CNOT
logical gate using atoms with a $\Sigma $-type three-level or a two-level
configuration in Cavity QED. In our scheme, we instead use the $\Lambda $%
-type three level configuration because it is rather difficult to utilize
the $\Sigma $-type level structure with SQUIDs.

\bibitem{s34}  A. S\o rensen and K. M\o lmer, Phys. Rev. Lett. {\bf 82,}
1971 (1999).

\bibitem{s35}  J. S. Bell, Physics (Long Island City, N. Y.) {\bf 1}, 195
(1965).

\bibitem{s36}  A. K. Ekert, Phys. Rev. Lett.{\bf \ 67}, 661 (1991).

\bibitem{s37}  D. Deutsch and R. Jozsa, Proc. R. Soc. London A {\bf 439},
553(1992).

\bibitem{s38}  C. H. Bennett, G. Brassard, C. Cr\'epeau, R. Jozsa, A. Peres,
and W. K. Wootters, Phys. Rev. Lett. {\bf 70,} 1895 (1993).

\bibitem{s39}  V. Buzek and M. Hillery, Phys. Rev. A {\bf 54}, 1844 (1996).

\bibitem{s40}  J. W. Pan, M. Daniell, S. Gasparoni, G. Weihs, and A.
Zeilinger, Phys. Rev. Lett. {\bf 86}, 4435 (2001).

\bibitem{s41}  C. A. Sackett, D. Kielpinski, B. E. King, C. Langer, V.
Meyer, C. J. Myatt, M. Rowe, Q. A. Turchette, W. M. Itano, D. J. Wineland
and C. Monroe, Nature {\bf 404}, 256 (2000).

\bibitem{s42}  E. Hagley, X. Maitre, G. Nogues, C. Wunderlich, M. Brune, J.
M. Raimond, and S. Haroche, Phys. Rev. Lett. {\bf 79}, 1 (1997); S. Osnaghi,
P. Bertet, A. Auffeves, P. Maioli, M. Brune, J. M. Raimond, and S. Haroche,
Phys. Rev. Lett. {\bf 87}, 037902 (2001).

\bibitem{s43}  L. X. Li and G. C. Guo, Phys. Rev. A {\bf 60}, 696 (1999).

\bibitem{s44}  M. A. Nielsen and I. L. Chuang, {\it Quantum Computation and
Quantum Information}, (Cambridge University Press, Cambridge, England, 2001).

\bibitem{s45}  G. Brassard, S. L. Braunstein, and R. Cleve, Physica D {\bf %
120}, 43 (1998).

\bibitem{s46}  M. Brune, E. Hagley, J. Dreyer, X. Ma\^\i tre, A. Maali, C.
Wunderlich, J. M. Raimond, and S. Haroche, Phys. Rev. Lett. {\bf 77}, 4887
(1996).
\end{references}
\end{document}